\documentclass[twocolumn,longauth]{aa}  

\usepackage{graphicx}
\usepackage{txfonts}
\usepackage{subcaption}
\usepackage{lscape}
\usepackage{placeins}

\usepackage{natbib}
\usepackage{hyperref}
\hypersetup{
   hidelinks=False,
   colorlinks=true,
   linkcolor=green,
   citecolor=blue,
   urlcolor=cyan,
}
\usepackage{cleveref}

\crefformat{figure}{Appendix~#2Fig.~#1#3}  
\crefformat{table}{Appendix~#2Tab.~#1#3}  
\crefformat{equation}{Appendix~#2Eq.~(#1)#3}  
\crefformat{section}{Appendix~#2Sec.~#1#3}  

\begin{document}

   \title{The ALMA-QUARKS Survey: Discovery of Superfine Dusty Fibrils inside Massive Star-forming Clumps}


   \author{Yan-Kun Zhang\inst{1,2}\thanks{yan-kun.zhang@lam.fr}        
        \and Tie Liu\inst{2}\thanks{liutie@shao.ac.cn}
        \and Wenyu Jiao\inst{2}\email{wenyujiao@shao.ac.cn}
        \and Pak-Shing Li\inst{2}\email{pakshingli@shao.ac.cn}
        \and Jia Zeng\inst{2,3}\email{zengjia@shao.ac.cn}
        \and Chao Zhang\inst{4,5}\email{zhangchao920610@126.com}
        \and Pablo García\inst{6,7}\email{astro.pablo.garcia@gmail.com}
        \and Mika Juvela\inst{8}\email{mika.juvela@helsinki.fi}
        \and Guido Garay\inst{6,9}\email{guido@das.uchile.cl}
        \and Amelia M. Stutz\inst{10}\email{amelia.stutz@gmail.com}
        \and Sami Dib\inst{11}\email{sami.dib@gmail.com}
        \and Dezhao Meng\inst{2,3,12}\email{mengdezhao@xao.ac.cn}
        \and Jian-Cheng Feng\inst{2}\email{fengjiancheng@shao.ac.cn}
        \and Dongting Yang\inst{2}\email{dongting@shao.ac.cn}
        \and Fengwei Xu\inst{11}\email{fengwei@mpia.de}
        \and Anandmayee Tej\inst{13}\email{tej@iist.ac.in}
        \and Enrique Vázquez-Semadeni\inst{14}\email{e.vazquez@irya.unam.mx}
        \and Gilberto C. Gómez\inst{14}\email{g.gomez@irya.unam.mx}
        \and Yong Zhang\inst{15,16}\email{zhangyong5@mail.sysu.edu.cn}
        \and Xindi Tang\inst{12}\email{tangxindi@xao.ac.cn}
        \and Paul F. Goldsmith\inst{17}\email{paul.f.goldsmith@jpl.nasa.gov}
        \and Kee-Tae Kim\inst{18}\email{ktkim@kasi.re.kr}
        \and James O. Chibueze\inst{19,20}\email{james.chibueze@gmail.com}
        \and Zhiyuan Ren\inst{21}\email{renzy@nao.cas.cn}
        \and Patricio Sanhueza\inst{22}\email{patosanhueza@gmail.com}
        \and Aiyuan Yang\inst{21,23}\email{yangay@bao.ac.cn}
        \and Jihye Hwang\inst{24,25}\email{hwang.jihye.514@m.kyushu-u.ac.jp}
        \and Shanghuo Li\inst{26,27}\email{shli@nju.edu.cn}
        \and Tapas Baug\inst{28}\email{tapas.polo@gmail.com}
        \and Shivani Gupta\inst{29,30}\email{shivani.gupta@iiap.res.in}
        \and Swagat R. Das\inst{31}\email{swagat@das.uchile.cl}
        \and Gang Wu\inst{12}\email{wug@xao.ac.cn}
        \and Jianjun Zhou\inst{12}\email{zhoujj@xao.ac.cn}
        \and Chang Won Lee\inst{32,33}\email{cwl@kasi.re.kr}
        \and Lokesh Dewangan\inst{34}\email{loku007@gmail.com}
        \and Prasanta Gorai\inst{35,36}\email{prasanta.astro@gmail.com}
        \and Tianning Lyu\inst{2,3}\email{lvtianning@shao.ac.cn}
        \and Lei Zhu\inst{6}\email{lzhupku@gmail.com}
        }

   \institute{Laboratoire d'Astrophysique de Marseille (UMR 6110 CNRS \& Université de Provence), 38 rue F. Joliot-Curie, 13388 Marseille Cedex 13, France
   \and State Key Laboratory of Radio Astronomy and Technology, Shanghai Astronomical Observatory, Chinese Academy of Sciences, 80 Nandan Road, Shanghai 200030, People's Republic of China
   \and University of Chinese Academy of Sciences, Beijing 100080, People's Republic of China
   \and Department of Physics, Taiyuan Normal University, Jinzhong, Shanxi 030619, People's Republic of China
   \and Institute of Computational and Applied Physics, Taiyuan Normal University, Jinzhong, Shanxi 030619, People's Republic of China
   \and Chinese Academy of Sciences South America Center for Astronomy, National Astronomical Observatories, Chinese Academy of Sciences, Beijing 100101, People's Republic of China
   \and Instituto de Astronomía, Universidad Católica del Norte, Av. Angamos 0610, Antofagasta, Chile
   \and Department of Physics, P.O.Box 64, FI-00014, University of Helsinki, Finland
   \and Departamento de Astronomía, Universidad de Chile, Camino el Observatorio 1515, Las Condes, Santiago, Chile
   \and Departamento de Astronomía, Universidad de Concepción,Casilla 160-C, Concepción, Chile
   \and Max Planck Institute for Astronomy, K\"{o}nigstuhl 17, 69117 Heidelberg, Germany
   \and State Key Laboratory of Radio Astronomy and Technology, Xinjiang Astronomical Observatory, Chinese Academy of Sciences, 150 Science 1-Street, Urumqi, Xinjiang 830011, People's Republic of China
   \and Indian Institute of Space Science and Technology, Thiruvananthapuram 695 547, Kerala, India
   \and Universidad Nacional Autónoma de México, Instituto de Radioastronomía y Astrofísica. Apdo. Postal 3-72, Morelia Mich. 58089, México
   \and School of Physics and Astronomy, SunYat-sen University, 2 Daxue Road, Tangjia, Zhuhai, Guangdong Province, People's Republic of China
   \and CSST Science Center for the Guangdong-Hongkong-Macau Greater Bay Area,SunYat-Sen University,Guangdong Province,People's Republic of China
   \and Jet Propulsion Laboratory, California Institute of Technology, 4800 Oak Grove Drive, Pasadnea CA 91109, USA
   \and Korea Astronomy and Space Science Institute, 776 Daedeokdae-ro, Yuseong-gu, Daejeon 34055, Republic of Korea
   \and UNISA Centre for Astrophysics and Space Sciences, College of Science, Engineering and Technology, University of South Africa, Cnr Christian de Wet Rd and Pioneer Avenue, Florida Park, 1709, Roodepoort, South Africa
   \and Department of Physics and Astronomy, Faculty of Physical Sciences, University of Nigeria, Carver Building, 1 University Road, Nsukka 410001, Nigeria
   \and National Astronomical Observatories, Chinese Academy of Sciences, A20 Datun Road, Chaoyang District, Beijing 100101, People's Republic of China
   \and Department of Astronomy, School of Science, The University of Tokyo, 7-3-1 Hongo, Bunkyo, Tokyo 113-0033, Japan
   \and Key Laboratory of Radio Astronomy and Technology, Chinese Academy of Sciences, A20 Datun Road, Chaoyang District, Beijing, 100101, People's Republic of China
   \and Institute for Advanced Study, Kyushu University, Japan
   \and Department of Earth and Planetary Sciences, Faculty of Science, Kyushu University, Nishi-ku, Fukuoka 819-0395, Japan
   \and School of Astronomy and Space Science, Nanjing University, 163 Xianlin Avenue, Nanjing 210023, People's Republic of China
   \and Key Laboratory of Modern Astronomy and Astrophysics (Nanjing University), Ministry of Education, Nanjing 210023, People's Republic of China
   \and S. N. Bose National Centre for Basic Sciences, Block-JD, Sector-III, Salt Lake City, Kolkata 700106, India
   \and Indian Institute of Astrophysics, Koramangala II Block, Bangalore 560034, India
   \and Pondicherry University, R.V. Nagar, Kalapet, 605014, Puducherry, India
   \and Departamento de Astronomía, Universidad de Chile, Camino el Observatorio 1515, Las Condes, Santiago, Chile
   \and Korea Astronomy and Space Science Institute, 776 Daedeokdae-ro, Yuseong-gu, Daejeon 34055, Republic of Korea
   \and University of Science and Technology, 217 Gajeong-ro, Yuseong-gu, Daejeon 34113, Korea
   \and Physical Research Laboratory, Navrangpura, Ahmedabad 380009, Gujarat, India
   \and Rosseland Centre for Solar Physics, University of Oslo, PO Box 1029 Blindern, 0315 Oslo, Norway
   \and Institute of Theoretical Astrophysics, University of Oslo, PO Box 1029 Blindern, 0315 Oslo, Norway}

   \date{Received xx xx, 20XX}

 
  \abstract{
   We report the discovery of more than 320 superfine dusty filamentary structures (fibrils) inside 121 massive star forming clumps that are located in widely different 
   Galactic environments (Galactocentric distances of $\sim1.2-12.9\ kpc$). These fibrils are identified from the 1.3~mm continuum emission in the ALMA-QUARKS survey, 
   which has a linear resolution of $\sim900$ AU for a source at $\sim3\ kpc$. The typical width of these fibrils identified in clumps is found to be $\sim0.01\ pc$, which 
   is about ten times narrower than that of dusty filaments in nearby clouds identified by the \textit{Herschel} Space Observatory. The mass ($M$) versus length ($l$) 
   relation for these fibrils follows $M\propto l^{2}$, similar to that of Galactic filaments identified in space (e.g., \textit{Herschel}) and ground-based single-dish 
   (e.g., \textit{APEX}) surveys. However, these fibrils in the ALMA-QUARKS survey are significantly denser ($\mathrm{N_{H_2} = 10^{23}-10^{24}\ cm^{-2}}$) than the 
   filaments found in previous \textit{Herschel} surveys ($\mathrm{N_{H_2} = 10^{20}-10^{23}\ cm^{-2}}$). This work contributes a large sample of newly found superfine 
   fibrils in massive clumps, following the identification of large 0.1$pc$-wide filaments and associated internal velocity coherent molecular fibers in nearby molecular 
   clouds, further emphasizing the crucial role played by filamentary structures in star formation at various physical scales.
   }


   \maketitle


\section{Introduction} \label{sec:Intro}

Interstellar filaments have a significant connection with the distribution and formation of dense cores and young stars 
(\citealt{2002ApJ...578..914H,2010A&A...518L.102A,2025RAA....25h5018J}). 
A notable population of filamentary structures in molecular clouds is formed by filament-hub systems, which feature a central hub that links various filaments 
\citet{2009ApJ...700.1609M}. Previous studies indicate that hub-filament systems (HFSs) are ubiquitous in proto-cluster clumps, playing a key role in high-mass star formation 
(e.g., \citealt{2010A&A...520A..49S,2014A&A...561A..83P,2020A&A...642A.177D,2020A&A...642A..87K,2022MNRAS.514.6038Z,2022ApJ...941...51H,2023MNRAS.520.3259X,
2023MNRAS.522.3719L,2023ApJ...953...40Y,2024A&A...689A..74A,2025A&A...696A.202S}). However, 
the internal structures of hub regions were not well studied in the past due to limited resolutions in previous observations.

Numerous continuum observational studies on nearby molecular clouds such as the IC 5146, the Aquila Rift and the Polaris cloud have shown that filaments have typical full 
width at half maximum (FWHM) of 0.1 $pc$ 
(e.g., \citealt{2011A&A...529L...6A,2014prpl.conf...27A,2022A&A...667L...1A,2025ApJ...984L..59A}). However, there are also instances 
of filamentary 
structures in star-forming regions with widths below 0.1 $pc$ 
(e.g., \citealt{2011ApJ...739L...2P,2016A&A...590A...2S,2017MNRAS.466.2529P,2019MNRAS.489..962H,2022A&A...657L..13P,2024ApJ...976..241Y,2026AJ....171...50H}). Based on analyses of the continuum 
emission in ALMA band-6, for example, \citet{2024ApJ...976..241Y} determined the filament width in Vela D cloud to be approximately 0.02 $pc$, and \citet{2026AJ....171...50H} 
even determined that width in G35.20-0.74N to be 0.015 $pc$.

In molecular line observations toward filaments, \citet{2013A&A...554A..55H} found fibers by applying the Friends In Velocity (FIVe) algorithm to identify 
velocity-coherent sub-structures within the L1495 filament, using the $\mathrm{C^{18}O(1-0)}$, $\mathrm{N_{2}H^{+}(1-0)}$ and $\mathrm{SO(J_{N}=3_2-2_1)}$ data obtained 
from the FCRAO 14-m telescopes. Utilizing ALMA observations, \citet{2018A&A...610A..77H} detected 0.035 $pc$-wide fibers within the Orion Integral Shape Filament in 
N$_2$H$^+$ (1-0) emission. \citet{2022ApJ...926..165L} also found $\sim0.04\ pc$ narrow filaments in ALMA observations with a similar spatial resolution toward another 
typical massive star formation region, NGC6334S, using H$^{13}$CO$^+$ (1-0) line. However, some works point out that measurements of 
filament widths in dense gas tracers such as $\mathrm{H^{13}CO^{+}}$ or $\mathrm{N_{2}H^{+}}$ are not very reliable since such observations are only probing gas material 
above a minimum critical/effective density \cite{2020MNRAS.499.3728P,2023A&A...672A.133S}.

In simulations, \citet{2014ApJ...791..124G} found filaments that are formed by transverse gravitational contraction. These filaments had Plummer-like radial profiles 
approaching a logarithmic slope of $-2$, extended out to $\sim1\ pc$ and density $\sim100$ cm$^{-3}$, and densities peaked at $\sim10^4$ cm$^{-3}$. Also, they showed that 
the gas flows along the filaments to the central hubs (see also in \citet{2011MNRAS.411.1354S}). However, \citet{2017MNRAS.467.4467S} observed filamentary structures with 
widths ranging from 0.01 to 0.1 $pc$. \citet{2022MNRAS.509.1494P} pointed out that the filament width of $\sim0.1\ pc$ requires for a circumstance with Mach numbers 
$\mathcal{M}\leq3$, if magnetic fields and trans-sonic turbulence are negligible.

It seems that the understanding of different kinds of interstelar filamentary structures (e.g., filaments and fibers) is still biased by different limited or specified 
samples and resolution effects in observations, and thus more careful examination are required. In particular, previous filament studies were mostly limited to the nearby 
clouds or low-mass star formation regions. In addition, how filamentary structures form and evolve in clustered, high-mass star-forming regions and their roles in high-mass 
star formation remains an open question \citep{2023ASPC..534..153H}.  With the unprecedented sensitivity and resolution provided by ALMA, it is now possible to observe and 
analyze the internal structures and physical properties of filaments at larger distances more distinctly, therefore probing more extreme and more massive environments. As 
massive star-forming regions are typically situated beyond 2 $kpc$ \cite{2009PASP..121..213C,2014MNRAS.443.1555U}, the enhanced resolution of ALMA will unveil previously 
unnoticed details related to high-mass star formation especially in filament-hub systems.

In this study, we utilized the ALMA-QUARKS (``Querying Underlying mechanisms of massive star formation with ALMA-Resolved gas Kinematics and Structures'') survey data 
offering superior angular resolution (0.3$\arcsec$) to extract and assess filamentary structures within an large sample of massive star-forming clumps located in widely 
different Galactic environments covering a large range of Galactocentric distances ($\sim$0.5-12.7 $kpc$, \citealt{2020MNRAS.496.2790L}), where \citet{2022MNRAS.514.6038Z} 
and \citet{2026arXiv260404501Z} found filaments are ubiquitous. This study focuses on the statistical properties (such as width, mass, and length) of filamentary structures 
in those regions, setting the foundation for future exploration of their formation mechanisms and evolutionary pathways, as well as their roles in high-mass star formation.
\section{Observations and Data} \label{sec:obs}

The dataset used in this study is from the ALMA-QUARKS survey (ALMA Project ID: 2021.1.00095.S; PIs: Lei Zhu, Guido Garay, and Tie Liu), which is conducted in Band 6 
(1.3~mm) by using the ACA (ALMA Compact Array) and 12-m arrays (C-2 and C-5 configurations) across 139 proto-clusters. The angular resolutions of the ACA and the 12-m 
C-2 and C-5 arrays are 4.5$^{\prime}$, 1.0$^{\prime}$, and 0.25$^{\prime}$, respectively. The receiver setup in the ALMA-QUARKS survey includes four spectral windows (SPWs), 
each spanning 1875 MHz, with a velocity resolution of approximately $\sim1.3$ km s$^{-1}$. Central frequencies of the four spectral windows are approximately 217.9, 220.3, 
231.4 and 233.5 GHz \cite{2024RAA....24b5009L}.

To reduce the data, the CASA (Common Astronomy Software Applications) software package (version: 6.6.0; \citealt{2007ASPC..376..127M}) was used. After data reduction, the 
typical root mean square (rms) levels for 1.3 mm continuum emission in the observational data obtained from the ACA array only, ACA plus 12-m C-2 arrays, and ACA plus 
both 12-m C-2 and C-5 arrays are 12, 0.6, and 0.2 mJy beam$^{-1}$, respectively. In this work, we utilize combined imaging data from all array configurations 
(ACA, 12 m C-2, and C-5), yielding an angular resolution of $\sim0.3^{\prime\prime}$, corresponding to linear resolutions ranging in 300 -- 3900 AU or 0.001 -- 0.019 $pc$ 
for sources with distances ranging in 1 -- 13 $kpc$. The maximum recoverable scale (MRS) is about 29$^{\prime\prime}$, which corresponds to linear scales of 0.09 -- 1.71 
$pc$ for sources at distances of 1 -- 13 $kpc$. The size of each pixel is $0.05^{\prime\prime}\times0.05^{\prime\prime}$. Additional details on data reduction can be found 
in \citet{2024RAA....24f5011X,2025ApJS..280...33Y,2024RAA....24b5009L}.
\section{Results and Discussion} \label{sec:results}
\subsection{Dusty fibrils identified in 1.3~mm continuum emission data} \label{subsec:Identificaion}

The massive star forming clumps detected in the 1.3~mm continuum emission within the ALMA-QUARKS sample contain ultra-compact ionized hydrogen (UC H{\sc II}) regions and/or 
massive young stellar objects (MYSOs), as exemplified in the left panel of Figure \ref{fig:width} for one source, namely I12320-6122. These clumps exhibit complex 
filamentary structures that are much shorter and narrower than the ones in nearby clouds identified in the \textit{Herschel} infrared continuum observations. Furthermore, 
they are even narrower than the velocity coherent fibers identified via molecular emissions in specific sources (see Section \ref{subsec:FittingWidth}). Therefore, we denote 
the filamentary structures revealed in the 1.3~mm continuum emission as ``dusty fibrils'', or ``fibrils'' for short.

The identification of fibrils was undertaken by utilizing the $FilFinder$ software \footnote{\url{https://github.com/e-koch/FilFinder}}. Considering the diverse nature 
(i.e., different evolutionary stages and various morphologies) of the ALMA-QUARKS sample (\citealt{2020MNRAS.496.2790L, 2024RAA....24f5011X}), as well as fluctuations in 
signal-to-noise ratios source-by-source, we employed a flexible parameter adjustment strategy to ensure optimal identification outcomes (see \cref{sec:Identification} for 
details). Through fine-tuning the key parameters in $FilFinder$, we successfully identified most ($>75\%$) filamentary structures within the ALMA-QUARKS sample that are 
consistent with visual inspection (see the atlas figures in \cref{sec:AppendixAtlas}). As the statistics of the column-density-defined contrast of fibrils by following 
\citet{2019A&A...621A..42A} presented in the right panel of \cref{fig:p-values}, most of the contrasts are larger than the criterion ($C^0\sim0.3$) used in 
\citet{2019A&A...621A..42A}, suggesting that our identification of fibrils is reliable. Note that we are not attempting to construct a complete sample of fibrils, but rather 
to identify the most prominent ones, i.e., elongated fibrils with strong dust emission.
\subsection{Widths of Fibrils} \label{subsec:FittingWidth}

\begin{figure*}[ht!]
\centering
\includegraphics[width=0.9\hsize]{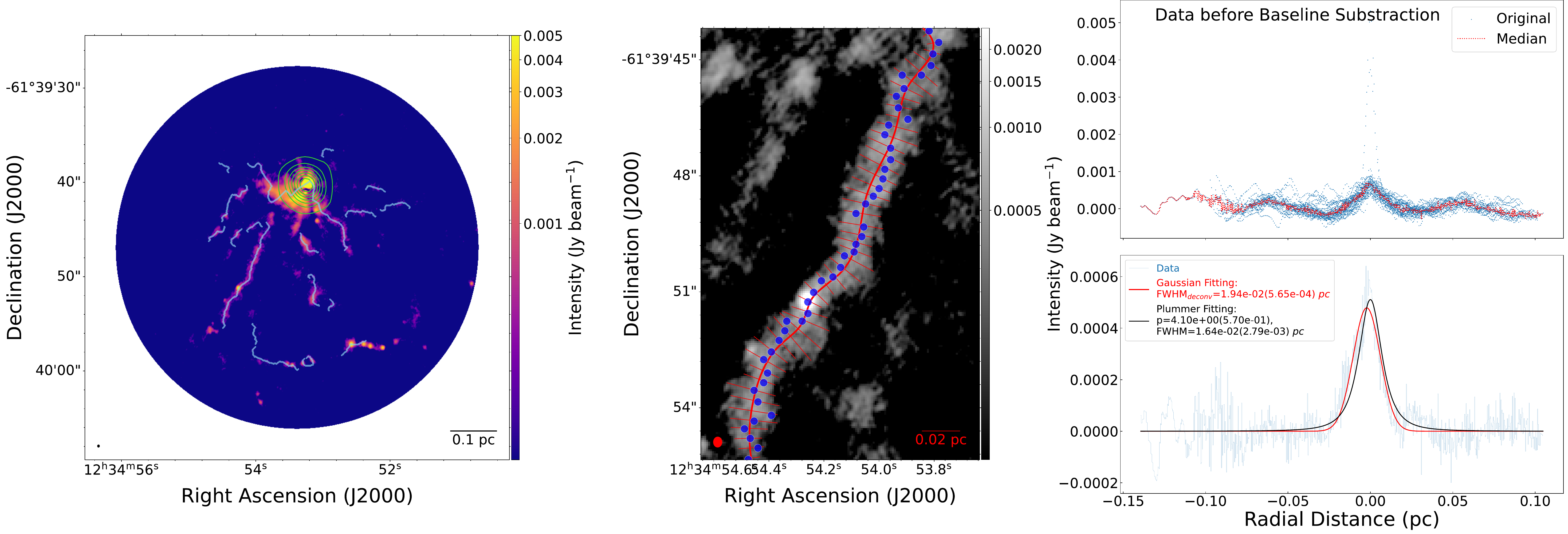}
\caption{(\textit{Left Panel}) The 1.3-mm continuum emission detected from the QUARKS survey in the I12320-6122 is shown as the background. The spines of fibrils are shown 
with sky-blue curves. The lime-green contours representing the H40$\alpha$ line emission detected in the `ALMA Three-millimeter Observations of Massive Star-forming 
regions' (ATOMS) survey (\citealt{2020MNRAS.496.2790L}), of which the start, step and end levels are 0.025, 1.520 and 13.500 Jy beam$^{-1}$ km s$^{-1}$, respectively.\quad
(\textit{Middle panel}) The grey background shows one of the fibrils identified in the I12320-6122. The thick red curve represent the spine, while the thin red lines are 
sampling paths where the profile data were extracted. The blue points mark locations where has the highest intensities along the sampling paths.\quad(\textit{Right Panel}) 
In the upper sub-panel, the blue points outline the profiles along the different thin red sampling paths, while the red points outline the averaged median profile in all 
these profiles. In the lower sub-panel, the blue curve presents the profile after deducting the baseline around the main peak by using \textit{Gildas/CLASS} from the 
averaged median profile, while the red and black curves present the Plummer and Gaussian fitting of this profile in \textit{SciPy}, respectively.}
\label{fig:width}
\end{figure*}

After extracting skeletons and masks for fibrils in each source with $FilFinder$, we at first employ $RadFil$\footnote{\url{https://github.com/catherinezucker/radfil}} to 
extract the radial intensity profiles of the most prominent fibrils with masked areas larger than 1000 square pixels. Sampling points are placed along fibrils' spines with 
a minimum separation of 5 pixels, as illustrated in the middle panel of Figure \ref{fig:width}. Since we found the built-in zeroth- and first-order linear baseline 
models in $RadFil$ lack the flexibility required to reliably subtract non-linear baselines or backgrounds, we then use 
$Gildas/CLASS$\footnote{\url{https://www.iram.fr/IRAMFR/GILDAS/}} to subtract the baselines, ensuring at the same time that the peaks and wings in intensity profiles were 
barely influenced. It should also be noted that the the median intensity profiles were used, because median profiles mitigate contamination from luminous dense cores and 
condensations located right to the spine of fibrils. Finally, We use $SciPy$ to compute Plummer and Gaussian full widths at half maximum (FWHM) for fibrils with aspect 
ratios greater than 10, following the formalism outlined in \cref{sec:WidthEquations}.

The parameters for 320 fibrils in 121 clumps, including lengths, Plummer FWHM, deconvolved Gaussian FWHM, and their respective errors, are listed in \cref{tab:cat}. In 
general, the widths derived from Plummer profile fitting align closely with those obtained from Gaussian profile fitting. It should be noted that the peak of the $p$-value 
distribution in the Plummer fit is approximately 5, which is larger than the $p=2$ value in the profile fittings of filaments identified in the \textit{Herschel} continuum 
data (see \cref{fig:p-values} and Section \ref{sec:discussion} for details).

\begin{figure}[ht!]
\centering
\includegraphics[width=0.6\hsize]{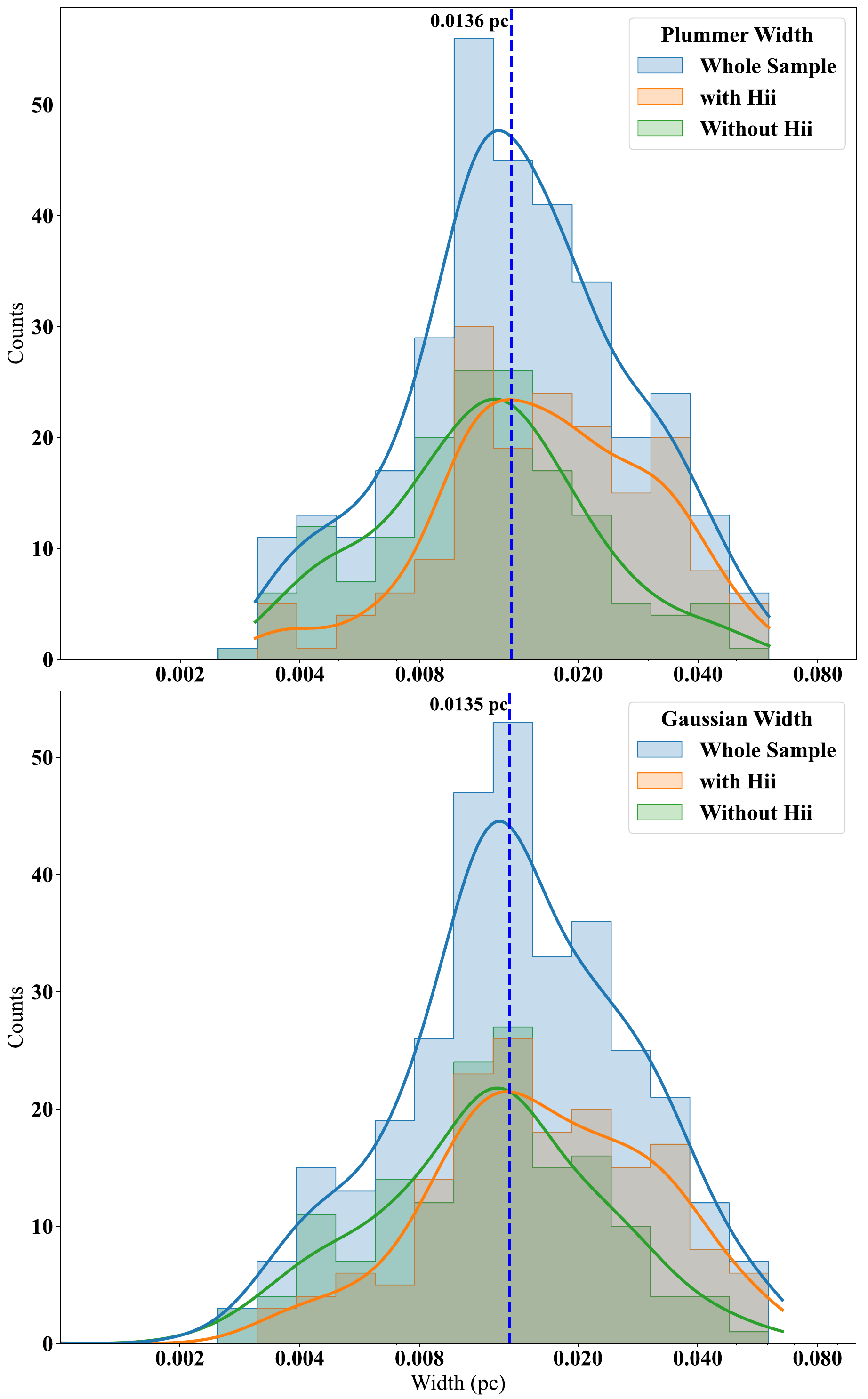}
\caption{\textit{Left Panel}: a histogram of the FWHM in the Plummer profile fitting for prominently long fibrils identified by using $FilFinder$. The orange, green and blue 
bins (curves) are the statistics (the Kernel Density Estimation (KDE)) for fibrils within sources containing UC H{\sc II} regions, without UC H{\sc II} regions traced by 
H40$\alpha$ emission, and for the whole sample, respectively. The dashed blue line marks the location of the median value in the whole sample.\quad\textit{Right Panel}: the 
same as the figure shown in the left panel, but for statistical results from the Gaussian profile fittings.
\label{fig:WidthResult}}
\end{figure}

The statistical analysis of fibril widths is presented in Figure \ref{fig:WidthResult}. Both the distribution of Plummer and Gaussian widths in this figure show a prominent 
peak at around 0.01 $pc$, and a secondary peak at approximately 0.03 $pc$ that is contributed mainly by fibrils adjacent to UC H{\sc II} regions. Besides, the fibrils in 
clumps associated with UC H{\sc II} regions (indicated by H40$\alpha$ emission) show statistically a little bit larger widths (by $\sim$0.002 $pc$) than that of fibrils 
without relation to UC H{\sc II} regions. The consistency of the fibril width obtained from these two distinct fitting methods suggests that fibrils derived from QUARKS 
high-resolution observations have a typical (median) width (${\rm FWHM}_{\rm typical}$) of $\sim$0.01 $pc$, which is about ten times narrower than \textit{Herschel} 
filaments (0.1 $pc$ in \citealt{2011A&A...529L...6A}) and even narrower than the $\sim0.04\ pc$-wide fibers mainly found in molecular line observations 
(\citealt{2018A&A...610A..77H}).

For the interferometric filtering effect, whereby extended or smooth emission can be partially or completely filtered out in interferometric observations, we refered to 
\citet{2024A&A...688A..30B} who investigated interferometric filtering effects for ALMA observations, and performed simulations of the ALMA-QUARKS observations using the 
\textit{simalma} tool within CASA, as presented in Appendix \cref{fig:simulations,fig:simulations4p5,fig:simulations4p2}. These figures demonstrate that neither the 
intensities nor the widths of the $0.01\,\mathrm{pc}$-wide fibrils reported here are underestimated in our simulations. Moreover, measurements of nearby 
$0.1\,\mathrm{pc}$-wide \textit{Herschel} filaments are also subject to only mild interferometric filtering biases at the typical large distances of the QUARKS sample.

\subsection{The Mass-Length Relation of Fibrils} \label{subsec:mass}

We expanded the area along the spine by the number of pixels corresponding to the Gaussian width on both sides and at both ends of the fibril, so as to calculated the 
physical area, total intensity $I_{\rm tot}$, total flux density $F_{\rm int}$ and mass $M$ of fibrils, following the Appendix \cref{eq:FluxDensity,eq:Mass}, respectively. 
In calculating the mass summarized in \cref{tab:cat}, the dust temperature is assumed to be equal to the clump averaged dust temperature obtained from the spectral energy 
distribution (SED) fit from the ATLASGAL and SIMBA samples (18 -- 46 K; see table A1 in \citealt{2020MNRAS.496.2790L}). Then, we calculate the typical H$_2$ column density of 
these fibrils as $\mathrm{10^{23}-10^{24}\ cm^{-2}}$. \cref{tab:cat} summarizes the masses of the primary fibrils, showing a mass range for these fibrils spanning 
from 0.5 to 4367.4 M$_{\odot}$, with median and mean values of 30 and 143 M$_{\odot}$, respectively. It should be noted that the masses of structures associated with 
H{\footnotesize II} regions may be overestimated due to contamination by free-free radiation, which will be further evaluated in future works.

\begin{figure*}[ht!]
\centering
\includegraphics[width=0.75\textwidth]{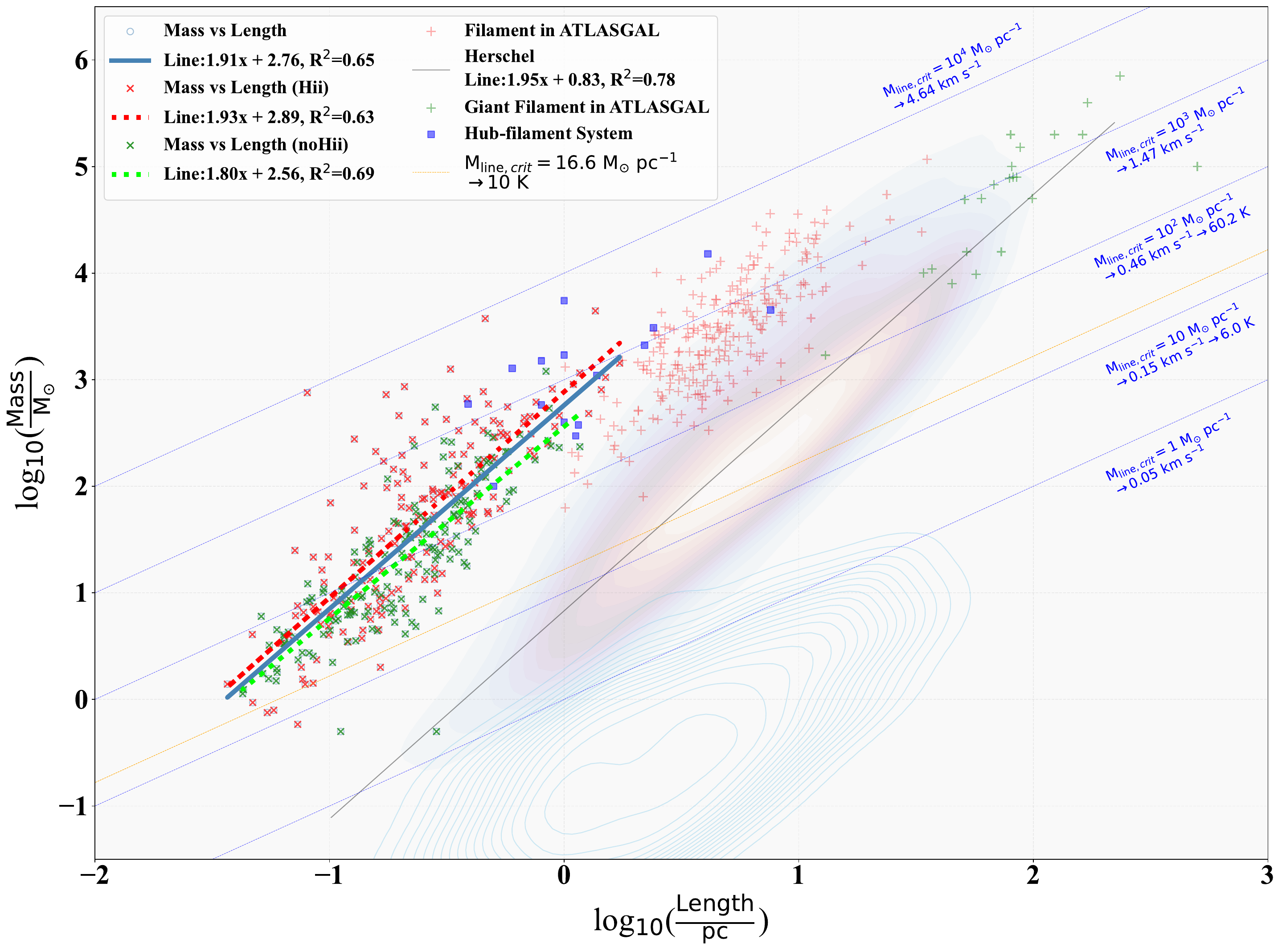}
\caption{Mass-Length relations among the H{\footnotesize I} filamentary clouds (\citealt{2014ApJ...789...82C,2023ASPC..534..153H,2026AJ....171...76P}), the dusty filaments 
(\citealt{2020MNRAS.492.5420S}), and the superfine fibrils. The colored and open contours are used to present the mass-length distributions of the H{\footnotesize I} clouds 
and the dusty filaments, respectively. The fitting results of their distributions are shown by the solid and sky-blue lines, correspondingly. The non-circled red and green 
crosses represent the common filaments and giant filaments detected in the ATLASGAL survey (\citealt{2016A&A...591A...5L}), respectively. In addition, the blue squares 
represent the hub-filament systems compiled in \citet{2025A&A...694A..69H}. For the QUARKS sample, we use dotted red and green lines to indicate the fitted mass-length 
relation of superfine fibrils within clumps/protoclusters that are and are not correspondingly found to emit H40$\alpha$ line emission, while the solid blue line are used to 
indicate the overall mass-length relation of superfine fibrils. The dotted blue lines are used to indicate the the critical line mass ($M_{{\rm line},crit}$) at varying 
temperatures and line widths.
\label{fig:MassLengthRelation}}
\end{figure*}

Obtaining the mass and length of fibrils, we present the mass-length relation of them in \ref{fig:MassLengthRelation}. We find from this figure that fibrils exhibit 
markedly distinct distributions in the mass-length space from neutral hydrogen (H{\footnotesize I}) filamentary clouds and from dusty filaments identified in 
\textit{Herschel} and ATLASGAL surveys. In general, Fibrils are much shorter and denser, with lengths extending from 0.04 to 1.72 $pc$ and masses from 0.5 to 4367.4 
M$_{\odot}$. Additionally, the distribution of fibrils partially overlaps with that of substructures in some HFSs in \citet{2025A&A...694A..69H}. 
Although fibrils and dusty filaments exhibit distinct distributions, their mass-length relationships show similar slopes ($M\propto l^{2}$). Since the majority 
of fibrils in this study lie above the critical line mass curve corresponding to $T=10$ K (see \ref{fig:MassLengthRelation}), they may likely fragment and collapse, unless 
turbulence and magnetic fields are taken into account to resist self-gravity of fibrils.

\section{Insights of Discovering Fibrils within Clumps} \label{sec:discussion}

The fibrils in this study differ from the filaments observed in the \textit{Herschel} data for two reasons. First, the peak of the $p$-value distribution in the Plummer fit 
is $\sim5$, whereas the $p$-value in the Plummer fit derived from the \textit{Herschel} data is $\sim2$ \cite{2019A&A...621A..42A}. 
Second, continuum observations from the QUARKS survey also show that these fibrils lack extended, gradually decreasing diffuse continuum emission. Previous studies 
on filamentary structures via molecular line emission by \citet{2022ApJ...935...57C} and \citet{2021ApJ...909...60S} used the VLA and GBT to study $\mathrm{NH_{3}(1,1)}$ 
emission in the Perseus B5 cloud and found that the width of filamentary structures in B5 was approximately 0.03 pc. Although \citet{2022ApJ...935...57C} found that ammonia 
molecular lines can trace low-density filamentary structures with a $p$-value of approximately 2, they also identified two other filamentary structures with $p$-values of 
3.1 and 3.5. In contrast, all of the filamentary structures identified by \citet{2021ApJ...909...60S} have $p$-values approximately equal to 3. These studies indirectly 
highlight the importance of the even narrower fibrils (typical 0.01-$pc$ width) with much larger $p$-values that we identified in the continuum images for star formation 
research. In addition, gaussian rather than Plummer function may be more suitable for describing the intensity profile of fibrils.

Our study indicates that superfine dusty fibrils are found to be ubiquitous in proto-clusters, and nearly all compact condensations or cores are located along them (see Figure 
\ref{fig:width} and \cref{fig:atlasExamples}). Furthermore, the width of fibrils (0.01 $pc$) are so comparable to the size of cores or condensations ($\leq0.01\ pc$; 
\citealt{2009ApJ...696..268Z}). These two evident clues further indicate that fibrils are closely related to the star formation process. It is seen in Figure \ref{fig:width} 
that some fibrils are clearly connected to already formed high-mass proto-stars or UC H{\footnotesize II} regions, indicating that they may play a crucial role in transporting 
gas from the natal gas clumps to the central high-mass proto-stars. Additionally, they could also fragment into dense condensations, which further give birth or feed mass to 
proto-stars \citep[e.g.,][]{2024ApJ...976..241Y,2025SciA...11.6953M}. Regardless of the crucial role these fibrils might play in the process of star formation, their 
formation process warrants further investigation.

\begin{figure*}[ht!]  
\includegraphics[width=1.0\textwidth]{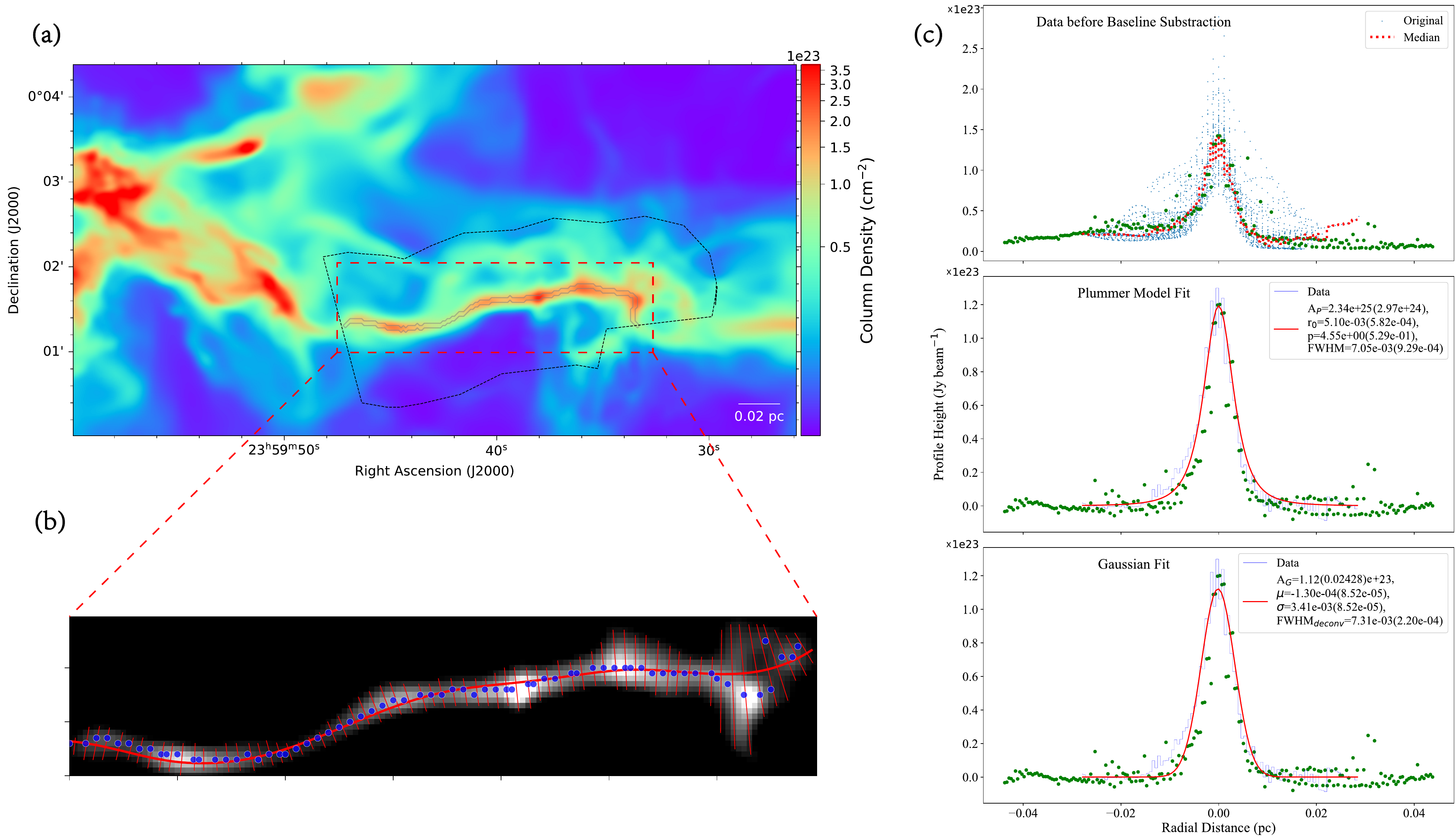}
\caption{Panel (a) presents the numerical simulation result. The colored background presents the column density distribution of the H$_2$. The gray curve marks the fibril.
\quad Panel (b) presents the identification result of the fibril marked by the grey curve in Panel (a), where the meaning of red curve/segments or blue points are the same 
as that shown in the middle panel of Figure \ref{fig:width}.\quad Panel (c) presents the Plummer and Gaussian fittings for the (column) density profile of the fibril marked 
in panels (a) and (b). In the upper subpanel, the blue points outline the profiles along the different lines perpendicular to the spine of the main fibril (i.e., the thin 
red segments shown in Panel (b)), while the red points outline the averaged median profile in all these profiles. The green datapoints present the averaged median curve for 
the data sampled from the expanding paths of the red short segments shown in panel (b) within the region sketched by the black dashed line in panel (a). In the middle 
subpanels, the blue curve and green datapoints present profiles that are deducted the baseline by using \textit{Gildas/CLASS} from the averaged median profile shown with 
blue profile and green datapoints in the upper subpanel, respectively. The the red curve presents the Plummer fitting by using \textit{SciPy}. In the lower subpanel, the 
meaning of all the curves and datapoints are the same as that shown in the middle subpanel, but for the Gaussian fitting.}
\label{fig:TestOnWidth_sim}
\end{figure*}

It is striking that the longest narrow fibrils in 121 out of 139 clumps have large aspect ratios ($>$10). To form these long fibrils, anisotropic agents such as shock 
compression, lateral gravitational contraction or magnetic fields are needed. In order to validate the existence and characteristic width of fibrils discovered in this work, 
we re-visit the simulations by \cite{2018MNRAS.473.4220L}, which simulated the formation of stellar clusters in a magnetized, filamentary infrared dark clouds at maximum 
$1.4\times10^{-4}$ $pc$ resolution utilizing the {\footnotesize ORION2} numerical simulation software (\citealt{2021JOSS....6.3771L}). A representative snapshot of this 
simulation is depicted in \ref{fig:TestOnWidth_sim}, illustrating the presence of high-density fibrils (average molecular hydrogen column densities up to 10$^{23}$ 
cm$^{-2}$). Employing the same methodology used for analyzing fibrils in the ALMA-QUARKS data but directly to the simulated column density data, specifically utilizing 
$FilFinder$ to identify skeletons, we measured the width of the most elongated fibril shown in Figure \ref{fig:TestOnWidth_sim}. This fibril has a length of approximately 
0.3 $pc$ and a Plummer FWHM value of roughly 0.007 $pc$. Here, we note that the rms of the simulation data is the smooth background or somewhat diffuse gas 
($\mathrm{\sim4.27\times10^{22}\ cm^{-2}}$), while the meaning of rms in the ALMA data is mainly the observational noises in the unit of $\mathrm{Jy\ beam^{-1}}$. It is 
interpreted that low-density filamentary structures first form through shock compression induced by supersonic turbulence, maintained by a moderately strong magnetic field. 
Then high-density fibril-like structures are formed through continuous merging and accretion regulated by global collapse (P. S. Li et al. 2026, submitted) and/or lateral 
gravitational contraction on the clump- or core-scale \citep{2014ApJ...791..124G}. Considering strong supersonic turbulence could be sustained by stellar feedback (outflows 
or H{\footnotesize II} regions) in the proto-clusters of the QUARKS sample \citep{2021MNRAS.507.4316B,2023MNRAS.520.3245Z,2024MNRAS.535.1364Z} and most clumps are 
gravitationally bound \citep{2016ApJ...829...59L,2020MNRAS.496.2821L}, we speculate that the fibrils identified in this work could be formed in a similar manner as in MHD 
simulations \citep{2018MNRAS.473.4220L}, which needs further tests in future observations.

\section{Summary} \label{sec:conclusion}

Based on the high-resolution ($\sim$0.3$^{\prime\prime}$) ALMA-QUARKS 1.3-mm dust continuum data, we have identified a population of very narrow filamentary structures with 
typical width of approximately 0.01 $pc$ in a large sample of massive star forming clumps. They appear much narrower and denser than conventional filaments or molecular 
fibers reported previously, and are linked to gas condensation formation and subsequent star formation process. We propose that these fibrils are likely formed by either 
shock compression induced by supersonic turbulence or transverse gravitational contraction and grow by continuous gas accretion regulated by global collapse, as observed in 
numerical simulations. The structure, dynamics and chemistry of these fibrils warrant further investigations.

\begin{acknowledgements}

This paper makes use of the following ALMA data: ADS/JAO.ALMA\#2021.1.00095.S. ALMA is a partnership of ESO (representing its member states), NSF (USA) and 
NINS (Japan), together with NRC (Canada), NSTC and ASIAA (Taiwan), and KASI (Republic of Korea), in cooperation with the Republic of Chile. The Joint ALMA Observatory is 
operated by ESO, AUI/NRAO and NAOJ.\\T.L. acknowledges the supports by the National Key R\&D Program of China (no. 2022YFA1603100), National Natural Science Foundation of 
China (NSFC) through grants no. 12073061 and no. 12122307, and the Tianchi Talent Program of Xinjiang Uygur Autonomous Region. This work was performed in part at the Jet 
Propulsion Laboratory, California Institute of Technology, under contract with the National Aeronautics and Space Administration (80NM0018D0004). P.S.L. acknowledges the 
supports by the National Key R\&D Program of China (No. 2022YFA1603101) and NSFC through grant No. 1241101426. Yan-Kun Zhang acknowledges the support from the Shanghai 
Post-doctoral Excellence Program (no. 2024689). MJ acknowledges the support of the Research Council of Finland Grant No.348342. G.G. and SRD gratefully acknowledge support 
by the ANID BASAL project FB210003. SRD also acknowledges support from the Fondecyt Postdoctoral fellowship (project code 3220162). G.C.G acknowledges support from 
UNAM-PAPIIT grant IN11082. S.L. acknowledges support from the National SKA Program of China with No. 2025SKA0140100, ``Double First-Class'' Funding with No. 14912217, and 
National Natural Science Foundation of China (NSFC) grant with No. 13004007. AS gratefully acknowledges support by the Fondecyt Regular (project code 1220610), and ANID 
BASAL project FB210003. C.W.L is supported by the Basic Science Research Program through the NRF funded by the Ministry of Education, Science and Technology (grant No. 
NRF-2019R1A2C1010851) and by the Korea Astronomy and Space Science Institute grant funded by the Korea government (MSIT; project No. 2025-1-841-02). PS was partially 
supported by a Grant-in-Aid for Scientific Research (KAKENHI Number JP23H01221) of JSPS.

\end{acknowledgements}


\bibliography{references}{}
\bibliographystyle{aa}

\begin{appendix}
\onecolumn
\section{Atlas of Superfine Fibrils} \label{sec:AppendixAtlas}
\setcounter{figure}{0}  

\begin{figure*}[ht]  
\includegraphics[width=1.0\textwidth]{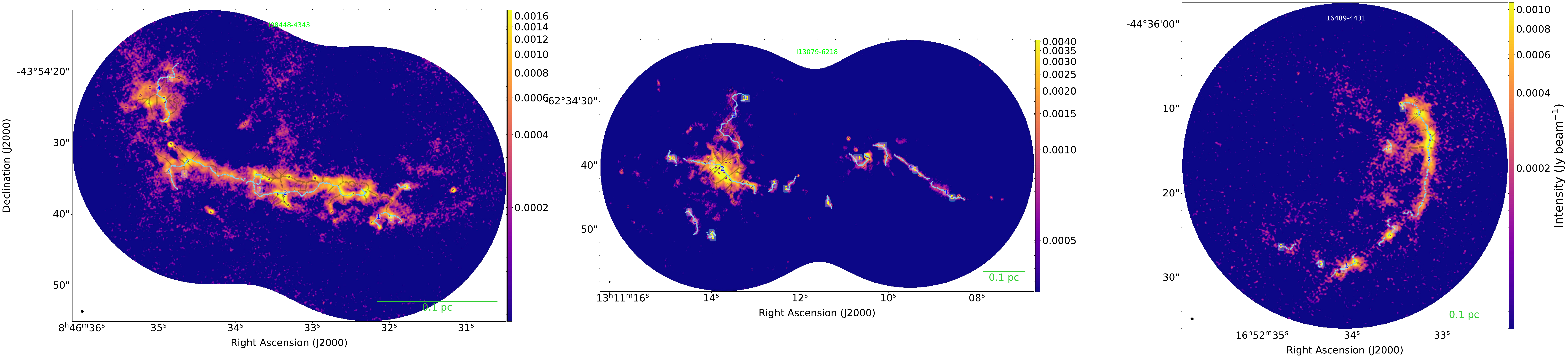}
\caption{Part of the atlas of the fibrils found in the QUARKS sample. The backgrounds show the 1.3-mm continuum emission detected in the 139 proto-clusters. The sky-blue 
and light-grey curves are used to present the spines and branches of the skeletons in fibrils, respectively. The blue and yellow numbers are used to correspondingly present 
the indexing of fibrils processed in $RadFil$ and $FilFinder$. In addition, we present the H40$\alpha$ emission detected in the ATOMS survey with limegreen contours, and 
cores found in Jiao et al. (in preparation) with purple circles.
\label{fig:atlasExamples}}
\end{figure*}

An atlas showcasing the superfine fibrils discovered in the QUARKS survey has been compiled, with a segment of it presented in \ref{fig:atlasExamples}. The full version of 
this atlas is available on Zenodo under an open-source Creative Commons Attribution license: \href{doi:10.5281/zenodo.20598299}{https://doi.org/10.5281/zenodo.20598299}, and 
it also can be found in the Github repository \footnote{https://github.com/yankunzhang0-creator/ALMA-QUARKS\_ATLAS-Fibrils/tree/main}.

\section{Method for Identifying Fibrils} \label{sec:Identification}

After testing a handful of sources that apparently have longer extended structures than other sources, we found that key parameters in \textit{FilFinder} include 
`flatten\_percent' within the `preprocess\_image' function, and `filt\_width', `size\_thresh' and `global\_thresh' within `create\_mask', which respectively aim to address 
essential structures, mask image margins, remove minor features, and establish noise thresholds. It detects fibrils by adopting a criterion that defines any standalone 
elongated entity, irrespective of whether the intensity is uniformly distributed, as a fibril if it harbors the longest fibrous structure, which is in line with the primary 
aim of emphasizing continuous structures without undue intricacy. In other words, the longest structure within an independent entity is recognized once as the `filament', 
but in this work as the `fibril', while adjacent structures are categorized as `branches'.

Given the sample diversity within the QUARKS survey (see Section \ref{sec:Identification}), a meticulous adjustment of the four $FilFinder$ parameters was undertaken to 
align with visual validation. Notably, the `filt\_width' was set at 10 pixels, and the `size\_thresh' at 400 square pixels. In cases of continuum images with the SNR $\geq$ 
3, the `flatten\_percent' was varied between 20 and 25. Conversely, for images with SNR $<$ 3, the `flatten\_percent' was chosen from 80, 85, 90, and 95. The 
`global\_thresh' in identifying all fibrils was set to be a factor of 2 of the rms in the observational images. Here, we regarded the intensity of the bright pixel in an 
image as the signal, and the rms of the image as the noise, because we found this is practical in measuring a lot of images in a survey.
\section{Fitting the Width of Fibrils} \label{sec:WidthEquations}

The equations for Gaussian and Plummer profiles are 
\begin{align}
    G(r) &= A_G \times e^{-\frac{(r - r_0)^2}{2\sigma^2}}, \label{eq:GaussianProfile} \\
    P(r) &= A_P \times \frac{R_{\rm flat}}{\left[1 + (\frac{r}{R_{\rm flat}})^2\right]^{\frac{p - 1}{2}}}, \label{eq:PlummerProfile}
\end{align}
where $r$ is the projected distance from the spline of the skeleton in a fibril, $r_0$ equals zero, $G(r)$ and $P(r)$ are the corresponding fitting intensity at different 
$r$ in the Gaussian and Plummer profiles, $R_{\rm flat}$ is the radius of the flat inner region of the fibril, $A_G$ and $A_P R_{\rm flat}$ are the corresponding fitted 
peak intensities of these two profiles. The $\sigma$ is the standard derivation, and $p$ is the power-law exponent at large radii ($r \gg R_{\rm flat}$) in the Plummer 
profile.

The FWHM (W) of the Gaussian and Plummer profiles are 
\begin{align}
    {\rm W}_{G} &= 2 \sigma \sqrt{2\times{\rm ln(2)}}, \label{eq:GaussianFWHM} \\
    \delta {\rm W}_{G} &= 2 \delta\sigma \sqrt{2\times{\rm ln(2)}}, \label{eq:GaussianFWHMerr} \\
    {\rm W}_{P} &= 2 R_{\rm flat}\sqrt{2^{\frac{2}{p - 1}} - 1}, \label{eq:PlummerFWHM} \\
    \delta {\rm W}_{P} &= \sqrt{\left(2 \times \delta R_{\rm flat} \sqrt{2^{\frac{2}{p-1}}-1}\right)^2 + \left[2\times{\rm ln(2)} \left(\sqrt{2^{\frac{2}{p-1}}-1}+1\right)\left(p-1\right)^{-2} \times R_{\rm flat} \times \delta p\right]^2}, \label{eq:PlummerFWHM_err}
\end{align}
where $\delta\sigma$, $\delta {\rm W}_{G}$, $\delta R_{\rm flat}$, $\delta p$ and $\delta {\rm W}_{P}$ are errors of $\sigma$, ${\rm W}_{G}$, $R_{\rm flat}$, $p$ and 
${\rm W}_{P}$, respectively.

The relation between W$_{G}$ and its corresponding de-convolved value W$_{Deconv,G}$ is 
\begin{align}
    {\rm W}_{Deconv,G} &= \sqrt{{\rm W}_{G}^2 - \theta_{phy}^2}, \label{eq:DeconvGaussianFWHM}
\end{align}
where $\theta_{phy}$ is the FWHM corresponding to the beam.

The error of the de-convolved value W$_{Deconv,G}$ is as the following
\begin{align}
    \delta {\rm W}_{Deconv,G} &= \frac{2 \times \delta\sigma \times {\rm W}_{G} \sqrt{2{\rm ln(2)}}}{{\rm W}_{Deconv,G}}, \label{eq:err4PlummerFWHM}
\end{align}
where the $\delta R_{\rm flat}$ and $\delta p$ are the error of the radius of the flat inner region of the fibril and the error of the power-law exponent at large radii in 
the Plummer profile fitting, respectively. The $\delta\sigma$ is the error of the standard derivation of fitted Gaussian profile.
\section{Statistics on the $p$-value in Plummer profile fittings and the column-density defined contrasts} \label{sec:p-values}
\setcounter{figure}{0}  

\begin{figure*}[htpb!]
\centering
\includegraphics[width=0.4\textwidth]{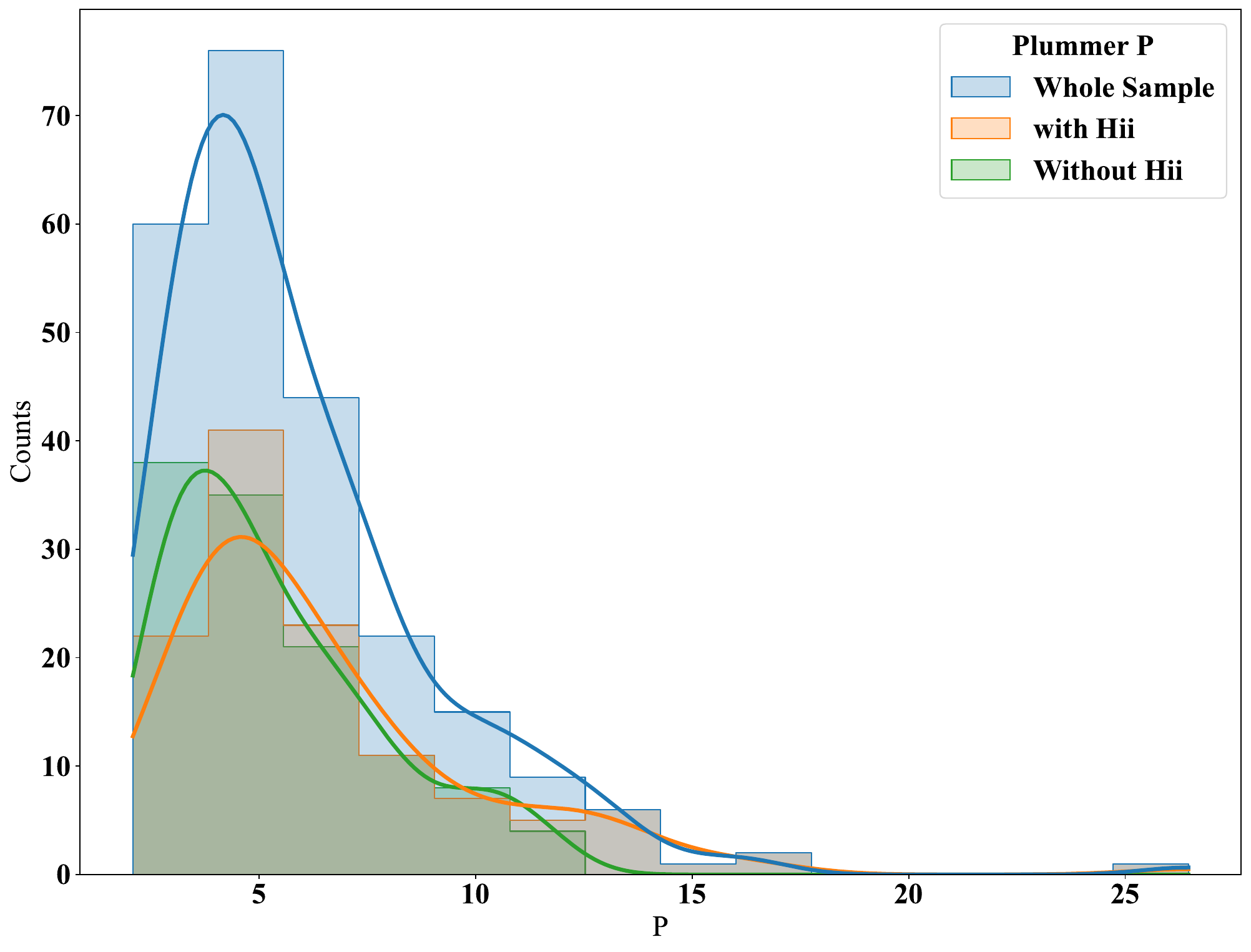}
\includegraphics[width=0.41\textwidth]{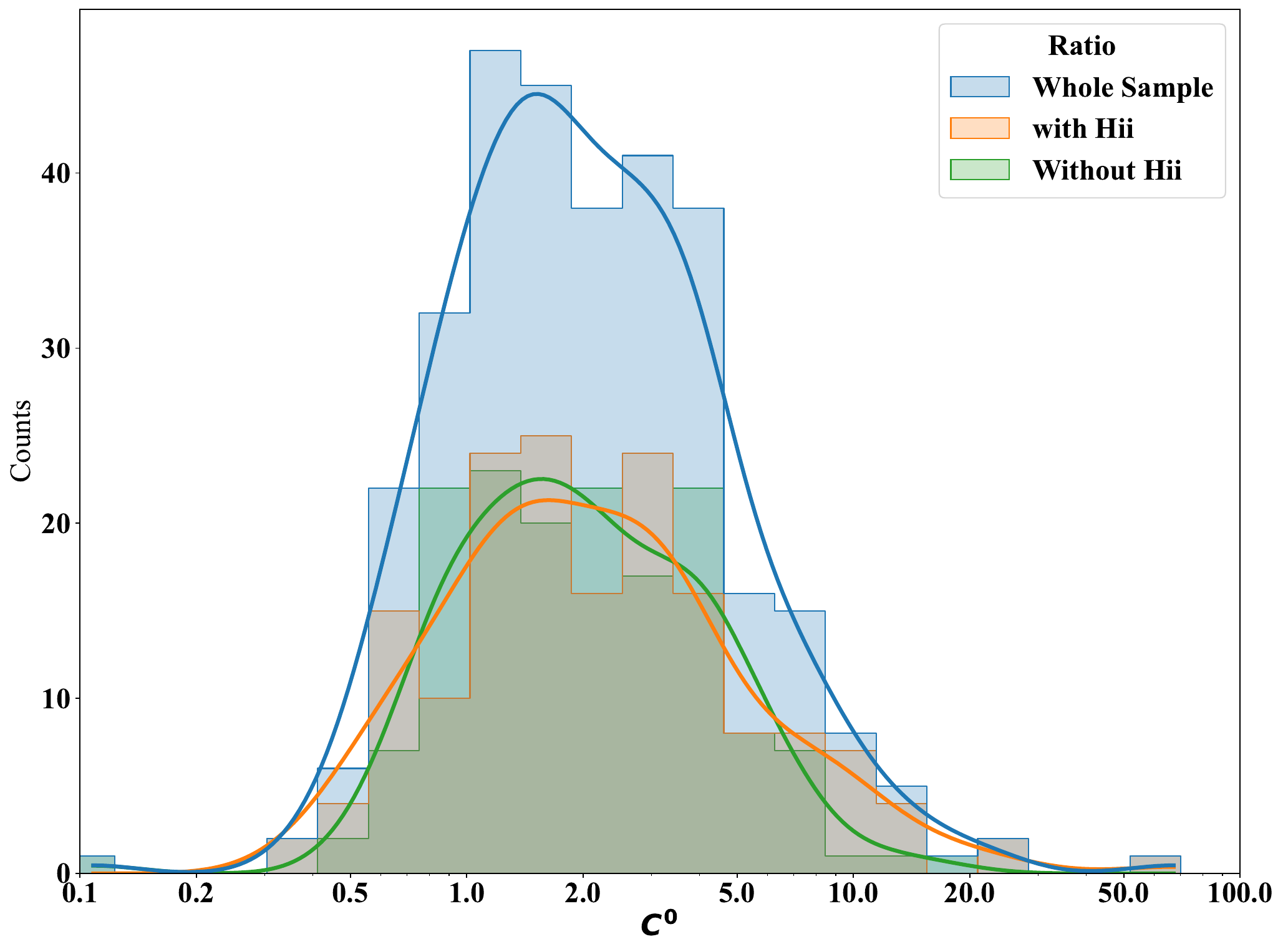}
\caption{\textit{Left panel}: The statistics of the $p$-value in the Plummer profile fittings. The orange, green and blue bins (curves) are the statistics (KDE) of 
$p$-values for fibrils within sources containing H{\sc II} regions, without H{\sc II} regions traced by H40$\alpha$ emission, and for the whole sample, respectively.\quad
\textit{Right panel}: The statistics of the column-density defined contrast for all reported fibrils in \cref{tab:cat}. 
\label{fig:p-values}}
\end{figure*}

The left panel of \cref{fig:p-values} present the statistical result of the $p$-values in the Plummer fittings of the intensity profiles of fibrils, where poor results are 
excluded if the error of $p$ is larger than itself. The right panel of \cref{fig:p-values} summarizes the column-density defined contrasts of fibrils to their surrounding 
backgrounds.

\section{Simulations on the Fibril Widths in ALMA observational Data} \label{sec:simulations}
\setcounter{figure}{0}  

\begin{figure*}[htpb!]
\centering
\includegraphics[width=1.0\textwidth]{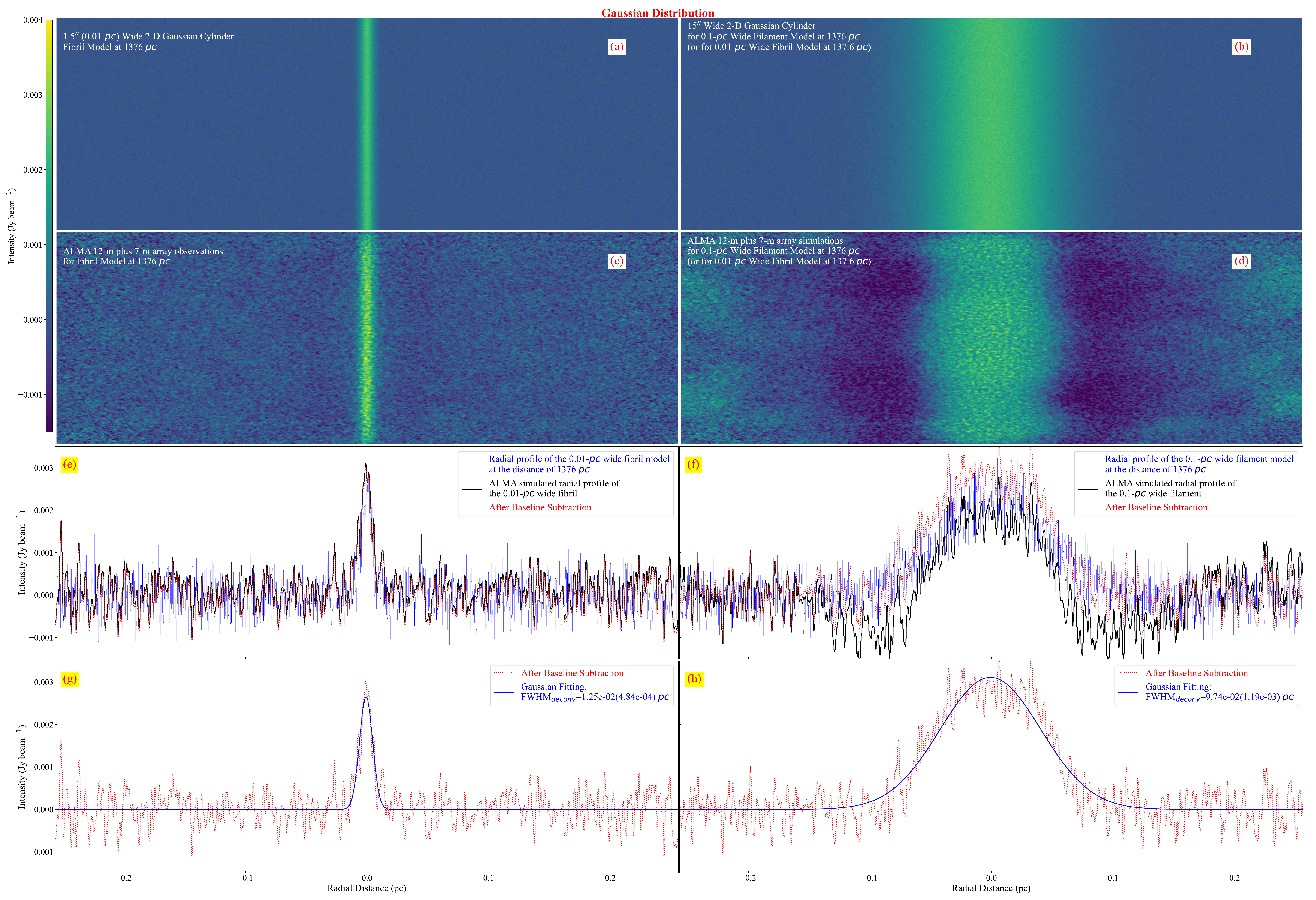}
\caption{Panel (a) shows the 0.01 $pc$-wide Gaussian cylinder fibril model located at the distance of $1376\ pc$. Panel (c) shows the simulation results toward the model 
shown in Panel (a), using the ALMA 12-m C2 and C5, plus ACA arrays. The blue and black curves in Panel (e) present the radial profiles of the model and the simulated 
fibrils, respectively. The dashed red curve present the result after the baseline subtraction of the simulated profile by using the $Gildas/CLASS$ tool. The blue 
curve in Panel (g) shows the Gaussian fit result of the intensity profile of the simulated fibril.\quad The meaning of components in Panels (b), (d), (f) and (h) are the 
same as that shown in Panels (a), (c), (e) and (g), respectively, but for the 0.1 $pc$-wide Gaussian cylinder filament.
\label{fig:simulations}}
\end{figure*}

\begin{figure*}[ht!]  
\centering
\includegraphics[width=1.0\textwidth]{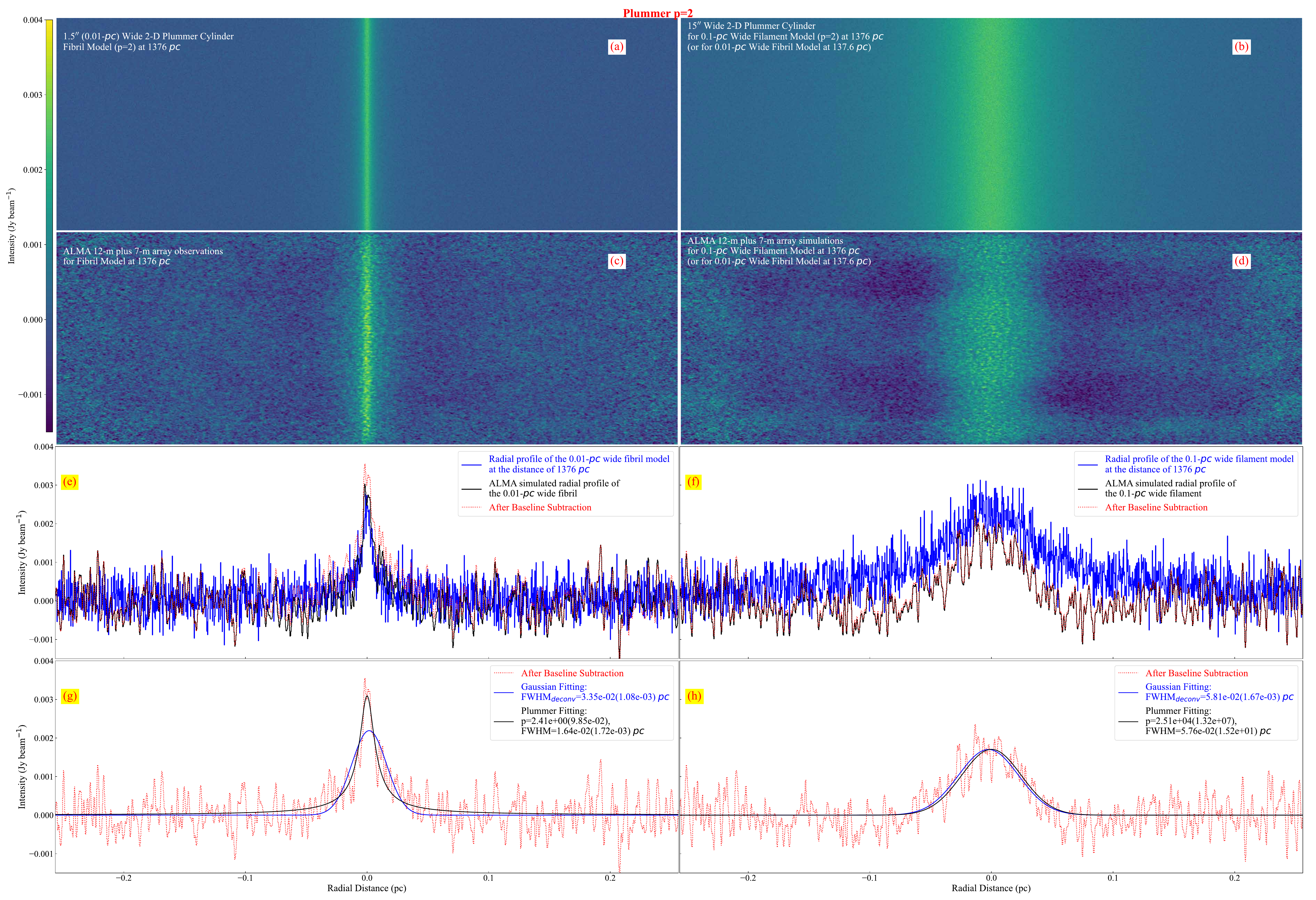}
\caption{The meaning of components in all the panels are the same as that shown in \cref{fig:simulations}, but for the 0.01 $pc$-wide fibril and the 0.1 $pc$-wide filament 
that have Plummer intensity profiles with the $p$=2. The additional black curves in Panel (g) and (h) present the Plummer fits to the intensity profiles that have been 
subtracted baselines.
\label{fig:simulations4p2}}
\end{figure*}

\begin{figure*}[ht]  
\centering
\includegraphics[width=1.0\textwidth]{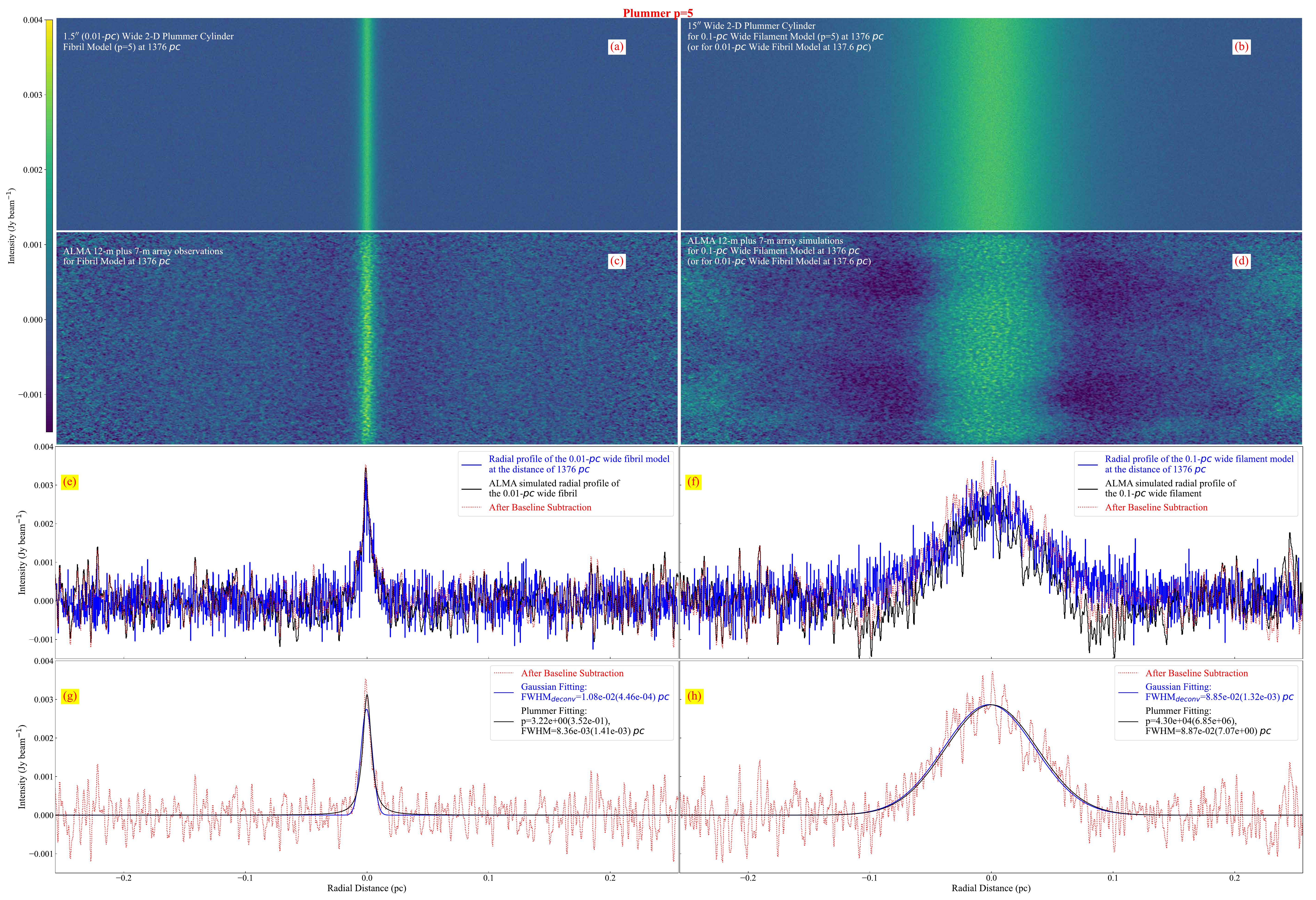}
\caption{The meaning of all the components in all panels are the same as that shown in \cref{fig:simulations4p2}, but for the fibril and the filament that have Plummer 
intensity profiles with the $p$=5.\label{fig:simulations4p5}}
\end{figure*}

We conduct simulations for one 0.01 $pc$-wide fibril and one 0.1 $pc$-wide \textit{Herschel} filament at the distance of $1376\ pc$, i.e., the distance that a 0.01 $pc$-wide 
fibril and a 0.1 $pc$-wide filament will have angular scales of $1^{\prime\prime}.5$ and 15$^{\prime\prime}$\footnote{One third about the FoV ($\sim45^{\prime\prime}$) for 
each source in the QUARKS sample.}, respectively. Assuming models are Gaussian cylinders, Plummer cylinders with the $p$=2 and 5 in the 2-D images, where the peak and RMS 
are corresponding to $3\times10^{-3}$ and $4\times10^{-4}$ Jy beam$^{-1}$ (i.e., similar continuum levels to the ALMA-QUARKS survey), and setting observations of using the 
ALMA ACA plus 12-m C2 and C5 arrays at 230 GHz, we got Appendix \cref{fig:simulations,fig:simulations4p2,fig:simulations4p5}. It should be noted that the radial profiles of 
the simulated observational fibril and filament shown in Panels (e) and (f) of Appendix \cref{fig:simulations,fig:simulations4p2,fig:simulations4p5} are sampled from the 
central part of the simulated images, because targets should be at the center of the FoVs and measurements from edges are not reliable. The negative values around the main 
object in the simulated images should be well avoided in a real observation and in the data reduction.

It is clear in these figures that intensities and widths of the fibrils are barely underestimated at the distance of $1376\ pc$ or even further, no matter what the model 
intensity profiles are. The wings around the peak in the simulated Plummer-shaped intensity profile of fibrils should also be detected and even not be suppressed under the 
configuration of using ALMA ACA plus 12-m C2 and C5 arrays at 230 GHz (see Appendix \cref{fig:simulations4p2,fig:simulations4p5}). The contrast to our observations that 
fibrils do not show clear wings around the peak in their intensity profiles, and both the Gaussian function and the Plummer function that many $p$ concern at 5 and some are 
even higher can well fit their intensity profiles, suggest that the intensity profile of fibrils are likely to be in the Gaussian form. Additionally, both the broaden width 
of the simulated fibril with the $p$=2 (see Panel (g) of \cref{fig:simulations4p2}), and the smaller $p$ of the fibril with its real $p$=5 (see Panel (g) of 
\cref{fig:simulations4p5}), suggest that the noise in the background could be picked into the measured signals, and thus contribute to wings of the fibril profiles. As for 
observations for the 0.1 $pc$-wide $Herschel$ filaments with distances less than 1376 $pc$, they will be influenced by the interferometry filtering effect, because the wings 
around the peak in their intensity profiles will be suppressed or lost, the observed intensity will be lower than their intrinsic intensity, and angular width will be larger 
than 15$^{\prime\prime}$ so that the observed widths will be underestimated about at least one third in the configuration of just using ALMA ACA plus 12-m C2 and C5 arrays to 
observe (See \cref{fig:simulations4p2} for detail). However, measurements of nearby 0.1 $pc$-wide Herschel filaments are subject to only mild interferometric filtering biases
at the typical large distances of the QUARKS sample.

In addition to the alignment with our observations, these simulations also suggest that widths of fibrils reported in this work will not be underestimated until they are 
located at distances less than 137.6 $pc$, which is a distance that the 0.01 $pc$-wide fibrils will have angular widths about 15$^{\prime\prime}$. However, this is not the 
case in our observations, because all fibrils in our sample are further than 1 $kpc$. We note that the possible existing finer structures inside the 0.1 $pc$-wide filament 
or even in the 0.01 $pc$-wide fibrils are not simulated, and are beyond the scope of this work.

\section{Masses of Fibrils} \label{sec:MassEquations}
\setcounter{equation}{0}

Given the total intensity $I_{\rm tot}$, the flux density $F_{\rm int}$ of each fibril can be calculated by using the following 
formulas\footnote{\url{https://science.nrao.edu/facilities/vla/proposing/TBconv}; \url{https://www.eaobservatory.org/jcmt/help/}}:
\begin{align}
    \left[\frac{F_{\rm int}}{Jy}\right] &= \frac{\left[\frac{I_{\rm tot}}{Jy\times{\rm pixels/beam}}\right]}{\left[\frac{\Omega}{\rm pixels/beam}\right]}, \label{eq:FluxDensity} \\
    \left[\frac{\Omega}{\rm pixels/beam}\right] &= \frac{\pi}{4\times{\rm ln}(2)}\left[\frac{\theta_{maj}}{pix}\right]\left[\frac{\theta_{min}}{pix}\right], \label{eq:Beam}
\end{align}
where $\Omega$ is the beam solid angle of the observation, and $\theta_{maj}$ and $\theta_{min}$ are the FWHM of the major and minor axes, respectively.

In addition to calculating the flux density of the fibrils, the dust temperature $T_{\rm dust}$ is also required to determine their mass $M$ by using the formula:
\begin{align}
    M = R\frac{F_{\rm int}D^{2}}{\kappa_{\nu}B_{\nu}(T_{\rm dust})}, \label{eq:Mass}
\end{align}
where $R$ is the gas-to-dust mass ratio (assumed to be 100), $D$ is the distance of the target from the Earth, $\kappa_{\nu}$ is the dust opacity at the frequency $\nu$, 
and $B_{\nu}(T_{\rm dust})$ is the Planck function at $T_{\rm dust}$ obtained from the SED fitting for clumps in the ATLASGAL and SIMBA 
observations (\citealt{2004A&A...426...97F,2018MNRAS.473.1059U,2020MNRAS.496.2790L}). In our calculation, $\kappa_{\nu}$ is assumed to be 0.899 cm$^{2}$ g$^{-1}$ at 
1.3-mm (\citealt{1994A&A...291..943O,2022A&A...657A..30T}).

Since the length and mass of fibrils were obtained, we calculated the critical line mass at different dust temperatures by using a hydrostatic and isothermal cylinder model 
(\citealt{1963AcA....13...30S,1964ApJ...140.1056O,2023ASPC..534..153H}):
\begin{align}
    \left(\frac{M}{L}\right)_{\rm critical} &= \frac{2c_{s}^{2}}{G} \sim 16.6\left(\frac{T}{10 {\rm K}}\right) {\rm M}_{\odot}\ {pc}^{-1}, \label{eq:LinearMass_critical}\\
    c_s &= \sqrt{\frac{k T_{\rm dust}}{\mu m_{\rm H}}}, \label{eq:SoundSpeed}
\end{align}
where $G$, $M$ and $L$ in Equation \ref{eq:SoundSpeed} are the gravitational constant, the mass and length of the fibril, respectively, and $k$, $\mu=2.37$, $m_{\rm H}$ and 
$c_s$ in Equation \ref{eq:SoundSpeed} are the Boltzmann constant, the mean molecular weight (\citealt{2008A&A...487..993K}), the mass of a Hydrogen atom and the sound speed, 
respectively.

However, the interstellar medium in reality exhibits not only thermal motion but also turbulence. Therefore, when calculating the critical line mass, both velocity 
broadening effects caused by these motions must be considered, as expressed by the formula 
(\citealt{2000MNRAS.311...85F,2014MNRAS.439.3275W,2023ASPC..534..153H}):
\begin{align}
    m_{vir}(\sigma) = \frac{2\sigma^{2}}{G} \sim 465\left(\frac{\sigma}{1\ {\rm km\ s^{-1}}}\right)^{2} {\rm M}_{\odot}\ {pc}^{-1}, \label{eq:LinearMass_sigma_critical}
\end{align}
where $\sigma$ and $m_{vir}(\sigma)$ denote the FWHM of a spectral line and its corresponding critical line mass, respectively.
\FloatBarrier 
\begin{sidewaystable*}
\section{Catalog of fibrils} \label{sec:AppendixTab}

The $FilFinder$ was utilized to detect filamentary structures across all 139 sources within the QUARKS sample. Subsequently, the mask and skeleton files obtained from 
this identification process were employed in conjunction with the 1.3~mm continuum image as input for $RadFil$ analysis, facilitating further Plummer and Gaussian fitting to 
determine the width of these filamentary structures. However, 18 sources were excluded from the analysis due to their complicated gas structures. The catalog details the 
identified fibrils within the remaining 121 sources, encompassing both basic metrics such as length and width, and derived parameters including mass (see \cref{tab:cat} for 
details).

\setlength{\tabcolsep}{2pt}
\setlength{\LTcapwidth}{1.2\textwidth}
\scriptsize{
\caption{Prameters Measured for Superfine Fibrils}
\label{tab:cat}
\centering
\begin{tabular}{cccccccccccccccccccccccc}
\hline\hline
\textbf{Name} & \textbf{ID(ff)} & \textbf{ID(rf)} &  \textbf{D}  &  \textbf{Td} &  \textbf{2$\sigma$} &  \textbf{L} & \textbf{pA} & \textbf{e\_pA} & \textbf{pP} & 
\textbf{e\_pP} & \textbf{pR} & \textbf{e\_pR} & \textbf{pW} & \textbf{e\_pW} & \textbf{gA} & \textbf{e\_gA} & \textbf{gW} & \textbf{e\_gW} & \textbf{Are} & \textbf{F} & 
\textbf{M} & \textbf{N$_{H_2}$} & \textbf{$C^0$}\\
\textbf{} & \textbf{} & \textbf{} & \textbf{($kpc$)} & \textbf{(K)} & \textbf{(U1\tablefootmark{a})} & \textbf{(pc)} & \textbf{(U2\tablefootmark{b})} & \textbf{(U2\tablefootmark{b})} & \textbf{} & \textbf{} & 
\textbf{($pc$)} & \textbf{($pc$)} & \textbf{($pc$)} & \textbf{($pc$)} & \textbf{(U2\tablefootmark{b})} & \textbf{(U2\tablefootmark{b})} & \textbf{($pc$)} & \textbf{($pc$)} & 
\textbf{($\mathrm{10^{-4}}$ $pc^{2}$)} & \textbf{(Jy)} & \textbf{(M$_{\odot}$)} & \textbf{(10$^{23}$ cm$^{-2}$)} & \textbf{}\\
\textbf{(1)} & \textbf{(2)} & \textbf{(3)} & \textbf{(4)} & \textbf{(5)} & \textbf{(6)} & \textbf{(7)} & \textbf{(8)} & \textbf{(9)} & \textbf{10} & \textbf{11} & 
\textbf{(12)} & \textbf{(13)} & \textbf{(14)} & \textbf{(15)} & \textbf{(16)} & \textbf{(17)} & \textbf{(18)} & \textbf{(19)} & \textbf{(20)} & \textbf{(21)} & 
\textbf{(22)} & \textbf{(23)} & \textbf{24}\\
\hline
...    &  ...  &  ...  &  ...    &  ...    &  ...    &  ...    &  ...    &  ...    &  ...    &  ...    &  ...    &  ...    &  ...    &  ...    &  ...    &  ...    &  ...    &  ...    &  ...    &  ...    &  ...    &  ...    &  ...    \\
I12320-6122  & 1     & 1     & 4.17  & 44.6  & 2.59  & 0.14  & 3.93  & 95.82  &  -      &  -      &  -      &  -      &  -      &  -      & 0.61  & 0.02  & 0.0147  & 0.0007  & 46.90  & 0.048  & 7.0   & 0.94  & 1.66  \\
-      & 5     & 2     &         &         &         & 0.29  & 47.18  & 7.91  & 4.08  & 0.56  & 0.0108  & 0.0016  & 0.0164  & 0.0028  & 0.48  & 0.01  & 0.0185  & 0.0006  & 113.80  & 0.120  & 17.4  & 0.96  & 1.70  \\
-      & 9     & 3     &         &         &         & 0.45  & 180.25  & 41.49  & 3.04  & 0.37  & 0.0052  & 0.0010  & 0.0103  & 0.0021  & 0.89  & 0.04  & 0.0101  & 0.0007  & 145.09  & 0.852  & 123.5  & 5.35  & 9.45  \\
I12326-6245  &  Hub    &         &         &         &         &         &         &         &         &         &         &         &         &         &         &         &         &         &         &         &         &         &    \\
...    &  ...  &  ...  &  ...    &  ...    &  ...    &  ...    &  ...    &  ...    &  ...    &  ...    &  ...    &  ...    &  ...    &  ...    &  ...    &  ...    &  ...    &  ...    &  ...    &  ...    &  ...    &  ...    &  ...    \\
\hline
\end{tabular}
\tablefoot{Column (1) lists the source name in the QUARKS sample. Columns (2) and (3) list the corresponding index of fibrils in $FilFinder$ and $Radfil$, respectively. 
However, there are also labels used to mark non-fibril sources in Column (2), where `hub' indicates a hub-dominated source, `sp' indicates a spiral-like sources, `st' 
indicates one apparent streamer infalling to its central source, `clu' indicates multiple sources residing in the proto-cluster, and `H{\tiny II}' indicates that the 
sources is dominated by H{\tiny II} regions, which are arc-, shell- or ring-shaped. Column (4) and (5) lists the adapted source distances and adopted dust temperatures, 
respectively. Column (6) lists mask thresholds, i.e., two times the noises of images. Column (7) lists the length of each fibril identified by using $FilFinder$ and 
measured by using $Radfil$. Columns (8)--(15) list the amplitude (pA), the error of the amplitude (e\_pA), the \textit{p}, the error of \textit{p}, the 
\textit{R$_{\rm flat}$} (pR), the error of \textit{R$_{\rm flat}$} (e\_pR), the FWHM (pW) and the error of the FWHM (e\_PW) in the Plummer fitting provided by $Radfil$, 
respectively. The cases that the errors of amplitudes are larger than amplitudes mean larger noises in the profile. The pR is not de-convolved. In addition, we removed the 
values of \textit{p} or the error of \textit{p} if either of them are larger than 100, no matter what the cause is: error propagation or bad Plummer fitting. Columns 
(16)--(19) list the amplitude, the error of amplitude, the \textit{FWHM} and the error of \textit{FWHM} from the Gaussian fitting in $Radfil$, respectively. The area, the 
flux, the mass, the column density of H$_2$, and the column-density defined contrast of fibrils are listed in Column (20)--(24), respectively.\\
\tablefoottext{a}{The tag `U1' represents `$\times10^{-4}$ Jy beam$^{-1}$'.}\\
\tablefoottext{b}{The tag `U2' represents `$\times10^{-3}$ Jy beam$^{-1}$'.}\\
The complete version of this table is machine-readable and on-line.}
}
\end{sidewaystable*}

\end{appendix}
\end{document}